\documentclass[3p]{elsarticle}

\usepackage{lineno,hyperref}

\modulolinenumbers[5]
\newcommand{\beq}{\begin{equation}}
\newcommand{\eeq}{\end{equation}}
\journal{Journal of \LaTeX\ Templates}

\usepackage{etoolbox}
\patchcmd{\abstract}{Abstract}{Summary}{}{}

\begin{document}

\begin{frontmatter}

\title{Necessity of 
ventilation for mitigating virus transmission  quantified simply
}


\author[mymainaddress,mysecondaryaddress]{Eric G. Blackman}\corref{mycorrespondingauthor}
\ead{blackman@pas.rochester.edu}

\author[mymainaddress]{Gourab Ghoshal}\corref{mycorrespondingauthor}
\cortext[mycorrespondingauthor]{Corresponding author}
\ead{gghoshal@pas.rochester.edu}

\address[mymainaddress]{Department of Physics and Astronomy, University of Rochester, Rochester, NY, 14627, USA}
\address[mysecondaryaddress]{Laboratory for Laser Energetics, University of Rochester, Rochester NY, 14623, USA}

\begin{abstract}

\noindent{\bf Background}
To mitigate the SARS-CoV-2 pandemic, officials have employed social distancing and stay-at-home measures,  with   increased attention to room ventilation emerging only more recently.  
Effective distancing practices for open spaces can be ineffective for poorly ventilated spaces, both of which are commonly filled with turbulent air.  This is typical for indoor spaces that use mixing ventilation.
While turbulence initially reduces the risk of infection near a virion-source, it eventually increases the exposure risk for all occupants in a space without ventilation.  
To complement detailed  models aimed at precision, minimalist  frameworks are useful
  to facilitate  order of magnitude estimates  for  how much ventilation provides safety, particularly when
  circumstances require practical decisions with limited options.
   
\noindent{\bf Method}
Applying  basic principles of transport and diffusion, we  estimate the time-scale for virions injected into a room of turbulent air to infect an occupant, distinguishing cases of low vs. high initial virion mass loads and virion-destroying vs. virion-reflecting walls. We consider the effect of an open window as a proxy for ventilation.

\noindent{\bf Findings}
When the airflow is dominated by isotropic turbulence, the minimum  area needed  to ensure safety depends only on the ratio of total viral load  to threshold load for infection.  

\noindent{\bf Interpretation}
The minimalist estimates  here  convey simply  that the equivalent of  ventilation by  
modest sized  open window in classrooms and workplaces significantly improves safety.
\end{abstract}

\begin{keyword}
\texttt{elsarticle.cls}\sep \LaTeX\sep Elsevier \sep template
\MSC[2010] 00-01\sep  99-00
\end{keyword}

\end{frontmatter}

\nolinenumbers

\section*{Introduction}

\label{s1}

The SARS CoV-2 virus, first reported in 
2019
~\cite{Wu_2020,Zhou_2020} 
 has since spread to at least 213 countries and territories  leading to an unprecedented global pandemic 
 \cite{Who}.
 The lack of therapeutics and vaccines have led public health officials to employ non-pharmaceutical interventions focused on  physical distancing measures and masks, but comparatively much
less on  ventilation. With  the resumption of
  visits to offices, bars, restaurants, salons and universities, where people are expected to be in close proximity in  for prolonged periods of time, ventilation and  HVAC filtering are  safety precautions that must be prioritized 
  ~\cite{Hadei_2020, Santarpia_2020}.  

Airborne viral particles like SARS-CoV-2, which cause Covid-19,  get trapped on moisture droplets that carry the virions~\cite{Qun_2020}.   Interior air is commonly, if not unavoidably,  turbulent. This  keeps 
 droplets airbone much longer than their free-fall time
\cite{Bourouiba2020}.  
The droplets form a spectrum of sizes, and those below $5\mu {\rm m}$ 
can remain airborne in aerosols for at least 3  hours \cite{vanDoremalen+2020}.
Since virions are attached to these  aerosol  droplets,    basic principles of  turbulent diffusion and  transport  or aerosols \cite{Batchelor1953,Csanady1973,Maxey1987,McComb1990,Blackadar+1997,Monin+2007a,Monin+2007b,Sreenivasan2019}
  become directly  applicable to guiding HVAC issues and  the   efficacy of masks \cite{Prather+2020}.
 Forced central air/heating in HVAC systems that use mixing ventilation 
 without sufficient replenishment of fresh air exacerbates the danger of 
 airborne virions. The  rapid spread of aerosol droplets in  room  of  turbulent air is exemplified    by spraying a scented aerosol   and measuring how quickly a person on the other side of the room can detect it. The benefits  of physical
 distancing 
are diminished by prolonged exposure to viral filled turbulent air in a closed room because virions are transported throughout 
by turbulent diffusion of the host droplets. Without a tightly sealed mask and proper eye protection, accumulated indoor-exposure is likely.   
 
Virions can also be transported by  HVAC systems between rooms. People staying home may be exposed to the virus in poorly ventilated apartment buildings with forced circulating air.  
Staying at home in isolated houses is not equivalent  to  staying home within a poorly ventilated apartment building.
 Evidence for such non-local transport of viruses  has been found in   restaurants in China~\cite{Li+2020,Lu2020}, a call-center in South Korea \cite{Park2020}, and a choir-setting in Washington state~\cite{Hamner2020}.
The   study of airbone disease transmission and ventilation  has a empirical and computational history \cite{Riley+1978,Beghein+2005,Aliabadi+2011,Smieszek+2019}.

Much about the  transmission modes of SARS CoV-2 remains unknown,   such as  the  viral load  required to cause infection~\cite{To_2020}.  Nevertheless, it is  crucial  to use basic principles that   we do know to  inform policy choices to reduce the risk of transmission~\cite{Hadei_2020}. As such, here we  use employ basic  concepts of turbulent transport to provide minimalist order-of-magnitude  estimates to quantify the efficacy of ventilation.

Specifically, we estimate   the  
  time scale for an individual to be infected in an enclosed space of arbitrary size, subjected to an injection of  virions.
We distinguish  between virion-absorbing vs. virion-reflecting  walls and cases in which the injected mass of virions is  sufficient  vs. insufficient   to infect a person over one diffusion  crossing time from the virial injector to the room boundary.  We  consider the role of an open window as a proxy for ventilation, and estimate the typical cross-section needed for safety.  We also discuss the role of a systematic drift velocity in the room.
Since the threshold viral load for severity of infection is unknown, despite the different loads produced by talking, breathing and other modes ~\cite{Balocco+2011,Buonanno+2020,Stadnytskyi+2020}, we present our results simply in terms of a dimensionless quantity--the virion mass required to  infect a single individual.  
Our results  demonstrate simply the importance of ventilation and filtering, 
complementing, but in agreement with, more detailed computational   \cite{Bourouiba2020,Li+2020,Lejeune+2020} 
and empirical efforts \cite{Somsen+2020}.

 
\begin{table}[t!]
\centering
\begin{tabular}{|l|l|r|l|}
\hline
 injected mass of virions & $M$    \\
\hline
mass of virions to infect one person& $M_c$    \\
\hline
$\#$ of people injected load can infect& $N=M/M_c$    \\
\hline
rate of virions encountering a face& ${\dot M}_h$    \\
\hline
density of virions uniformly spread over room& $\rho$    \\
\hline
room radius & $R$    \\
\hline
eddy size & $l$   \\
\hline
turbulent eddy  air speed & $v_l$    \\
\hline
eddy turnover time & $t_{l}=l/v_l$    \\
\hline
eddy diffusivity & $\nu_T=v_l l/3$    \\
\hline
 diffusion speed over scale $x\ge l$ & $v_{\textrm{dif}}(x)=\nu_T/x$    \\
  \hline
mean  air flow  & $v_{p}$    \\
\hline
face width & $h$    \\
\hline
face area & $A_h=h^2$   \\
\hline
required time for infection & $t_c $  \\
\hline
window area& $W $   \\
\hline
time for load to diffuse via window & $t_{\textrm{saf}}$   
 \\
\hline
\end{tabular}
\caption{\label{tab:5/tc}
list of variables 
}
\label{table1}
\end{table}
\bigskip

\noindent\fbox{%
    \parbox{\textwidth}{%
       {\bf Research in Context}
       
   \subsection*{Evidence before this study}    
       Coughing, sneezing and breathing release moisture droplets onto which 
       SARS-CoV-2 bind.  Droplets of size below $5\mu m$ in aerosols  remain viable and airborne with a half life of   three hours \cite{vanDoremalen+2020}.  Interior air is commonly turbulent and  diffuses aerosols throughout a room and building over much shorter time scales,  spreading the infectious virions over distances much larger than the 6ft recommended
       physical distancing separation. 
      \subsection*{Added value of this study}
We   quantify  the   mitigating role of ventilation  using simple  from simple  considerations of fluid
transport and diffusion, by  comparing the load of virions encountering a persons face in a room with and without   a window.  Specifically, this becomes a  determination of the size of the window or filter needed to maintain safety.  The results  help guide  practical  decisions  about safety and interior ventilation in spaces where people spend extended periods, such as classrooms and office buildings.
\subsection*{Implications of all the available evidence}
Proper ventilation mitigates the airborne spread of SARS-CoV-2 and needs to  be viewed along with masks and physical distancing as the third pillar of prevention,  We show its efficacy quantitatively in a simple broadly accessible   way.}%
}
\section*{Results}
\label{s2}

While progressively
  smaller droplets take longer and longer to fall out of the air \citep{Somsen+2020}
  after the initial load (cough, breathing, or sneeze), we are mainly concerned 
  with  particles  that remain suspended in the air after it is well mixed, namely
 these  $\le 5 \mu$ that can remain airborne for $3$hr \citep{vanDoremalen+2020}.
  Such particles  are  equivalent carriers of virions for present purposes.  
   
We assume that  a combined mass $M$ of  such viron-loaded aerosol droplets (VLADs) 
are
 injected  
in a room of  radius  $R$, and that the room  has steady  turbulent air of  local eddy air speed $v_l$ and
 eddy scale $l<<R$.    
We define  $t_c$ as the time scale for a single person to get infected   
and $M_c$ is critical mass of VLADs needed to infect one person. 
These two quantities are related by 
\beq
t_c= M_c/ {\dot M}_h,
 \label{1}
\eeq
where  ${\dot M}_h$ is the rate of  air mass encountering  a human face of diameter $h$.
We assume that the  threshold viral load  for infection is small compared to the total viral content in the room.
In Table ~\ref{table1}, we list the  variables used in the following calculation results.


\subsection*{Large viral injection mass, no ventilation} 
\label{case1}

We first consider the case of a closed room (no ventilation), for which 
the injected mass load is sufficiently high that 
a single passing of the diffusion front after the viral injection 
is sufficient to cause infection.
For this case,
the time scale to infect any individual  is bounded by the diffusion time from the injection point to the room boundary. Assuming isotropic turbulence, the eddy diffusivity is $\nu_T = v_l \times l/3$. The critical time for infection is therefore      
\begin{equation}
\begin{array}{l}
 t_c \le R^2/\nu_T = 3 t_l \left(\frac{R}{l}\right)^2 \\
 \le 1.67\left({t_{l}\over 0.75{\rm s}}\right)\left({R\over 5{\rm m}}\right)^2\left({l\over 0.75{\rm m}}\right)^{-2}
 {\rm min}.
\end{array}
 \label{3a}
\end{equation}
Here $t_l = l/v_l$ is the eddy turnover time and its scaling of $0.75{\rm s}$ was estimated using an eddy scale of $l=0.75$m corresponding to injection  by a room fan at a flow speed $1 {\rm m/s}$.

\subsection*{Small viral injection mass, no ventilation} 
\label{case2}
The likely more commonly  important case is when  the initial injection of  viral mass is small enough so that the room is fully mixed with VLADs before  any individual is infected.   
First,  we consider two sub-cases without ventilation:  (i) completely absorbing walls (say if the walls are infused with anti-viral material such as copper fibers) and (ii) completely reflecting walls.
For the  former, sub-case virions are removed upon contact with the wall and no one in the room is infected.  
 
 For the  latter  sub-case, the  virion-carrying droplets remain airborne.
Then, given that the accumulated mass of virions on a single person is
\beq
{\dot M}_h \simeq \rho  v_{dif} A_h,
 \label{2}
 \eeq
 where 
\beq
\rho ={3M\over 4\pi R^3},
 \label{2a}
 \eeq 
and $v_{\rm dif}$ is the diffusion speed, which is simply the turbulent diffusion coefficient  $\nu_T$ divided by the net  displacement from the source to point of measurement.
Given that we are in the well-mixed regime, the relevant displacement is just the eddy
scale, since any new virion-mass accumulates via neighboring eddies. Therefore,
  \beq
 v_{\rm dif} =   {1\over 3} v_l , 
 \label{3}
 \eeq
Combining Eqns.~(\ref{2}),  (\ref{2a}), and (\ref{3})  into Eq.~(\ref{1}) yields, for the critical infection time,
\beq
  \begin{array}{l}
t_c=
13.11 \left({t_{l}\over 0.75\ {\rm s}}\right) \left({M/M_c\over 50}\right)^{-1}
\left({R \over 5{\rm m}}\right)^3\left({l\over 0.75{\rm m}}\right)^{-1}
 \left({h \over 0.2{\rm m}}\right)^{-2}
 {\rm min}, 
 \end{array}
 \label{5}
 \eeq
 where we have used fiducial values  to  facilitate  estimates. 
  Note, that while $M_c$ is unknown, in all cases it enters as as fraction of the input  viral mass. In the upper panel of 
 Fig. 1
 we plot Eq.~(\ref{5}) as a function of the room radius $R$ and injected load $N = M/M_c$. By definition, to prevent infection an occupant must spend a duration $t \leq t_c$  in the room. 
 
\subsection*{Role of an open window} 

To quantify the effect of ventilation, we consider  window of open area
 $W<<R^2$, that allows exchange of  interior and exterior air.
If pressure and temperature equilibrate everywhere,  turbulence  facilitates mixing 
interior VLAD filled air with virus-free air. 
VLADs leave the window with a mass loss rate
\beq
{\dot M}_{W}=\rho v_{\rm dif}W. 
\label{10}
\eeq
From the time  of VLAD  injection in the room, it  takes a single diffusion time from the source to the window
for  exterior air to interact with the viral air.  This has no effect for the case when the viral mass exceeds the critical threshold for a single infection.
 Nor does it have any bearing for absorbing walls, 
where the viral load is absorbed over one diffusion crossing time.

The ventilation is  however, very influential for the  reflecting wall case
of Eq.~({\ref 5}), which now requires modification.
For this case, the  VLAD  mass lost over a  time   $t >> R^2/ \nu_T
$  can be estimated as  $M_L (t) = {\dot M}_{W} t$. Dividing this  by $M$ and using 
equations  (\ref{2a}) and (\ref{10})
 gives the fraction of  VLAD mass lost from the room  as
  \beq
M_L(t)/M  \simeq  {1 \over 4 \pi} \left({v_lt\over R}\right)\left({W \over R^2}\right).
\label{11}
\eeq
Safety would  be achieved once $M_L \ge M(1-M_c/M)$, corresponding to 
$\ge 99\%$  replacement of the  room  air when $M_c
le 0.01 M$. (If using a ventilation system measured 
in conventional ventilation units of air changes per hour (ACH) \cite{Bearg1993}, for
$M_c/M < 0.01$ this would require $> 10 ACH$ to achieve this replacement in 1 hour
 since each ACH replaces only $63.2\%$
of the air.)

Solving  Eq.~(\ref{11})  for
the associated $t=t_{\rm saf}$ gives,
 \beq
t_{\rm saf} 
= {4\pi R^3\over v_l W}.
 \label{12a}
 \eeq
  Now we set Eq.~(\ref{5})
equal to Eq.~(\ref{12a}) to get the minimum open window area  $W_c$ ensuring $t_{saf}< t_c$, so that enough virions are removed before anyone gets infected.
This  gives 
\beq\begin{array}{l}
W_c\ge  2 \left( {h\over 0.2 {\rm m}}\right)^2 \left({M/M_c\over 50}\right) {\rm m^2}.
   \end{array}
 \label{12}
 \eeq
This is independent of the room size because the same viral mass in a larger room reduces the viral density. Therefore the flux of mass incident on a face, as well as on the surface of a window is reduced by the same factor. In the lower panel of Fig. 1b
we show $t_c/t_{\rm saf}$ as a function of $W$ and $N$. The intersection of the curved surface with the  plane is equation  (\ref{12}).

Given that $\rho$ is uniform  in the room for the fully mixed case by   diffusive action of  turbulence, the ratio of  mass flux through a window to that encountering a person's face is  approximately $W/h^2$.  Increasing this ratio enhances the probability for viral mass encountering the window.  Our  example shows that the strategy for safety is to  increase this ratio such that  the leftover virial density in the room is insufficient to produce  a critical infectious load for individuals during their period of stay. 

What is the role of masks in the context of the above estimates?  They   reduce
the surface area of the face by at least a factor of 2   since droplets must slide around the smaller open area
around the edges of the mask.   A factor of $\ge 2$ reduction in $h^2$ has the equivalent effect of reducing the required
size of the window area for safety by that same factor as seen from  Eq. \ref{12}.

A generalization of interest is the case in which there is a systematic bulk flow in the from one side to the other, for example from a pressure gradient or fan.  In this case,  the ratio of the flux through the window to that onto a face  depends on the angles of the bulk flow to the window, $\theta_w$ and  to the face, $\theta_f$. If the bulk flow is aimed toward the window, then only the latter angle enters and  this ratio of fluxes becomes $(v_T + v_P) W\over (v_T - v_P \cos \theta_f )h^2$ where $\cos \theta_f$ is the angle between the flow and the normal to the face. If  $|v_P \cos \theta_f | >> v_T$, then this ratio becomes simply $-W/(h^2 \cos \theta_f)$ highlighting  that simply having ones back to the flow is the safer orientation, as the flux ratio is negative when the flow passes from back to front.   In practice detailed modeling of flow around the head 
must be considered to assess the influence of boundary layers.

 \begin{figure}[t!]
\begin{center}
\includegraphics[width=0.45\textwidth]{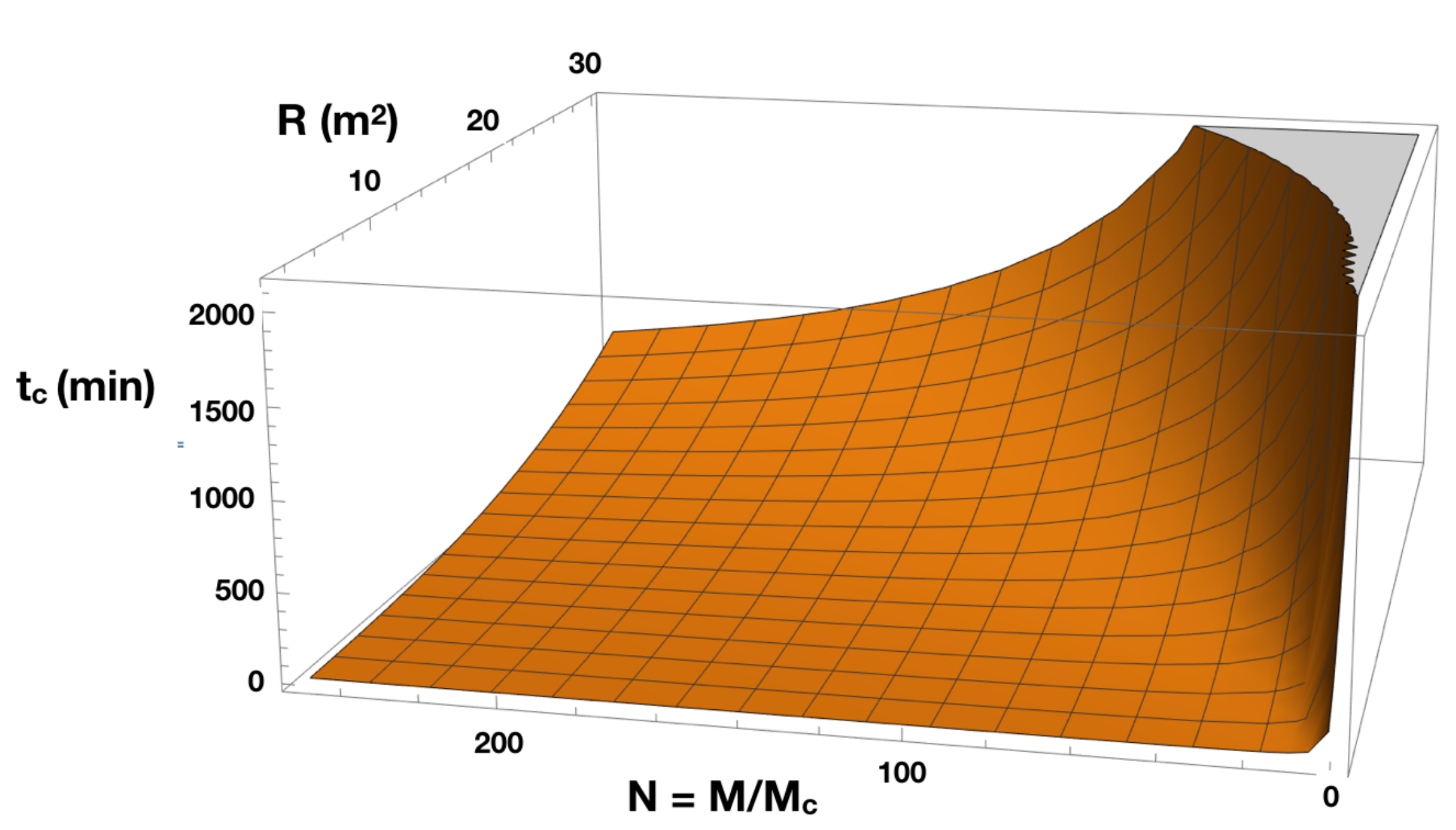}
\vskip0.1in
\includegraphics[width=0.45\textwidth]{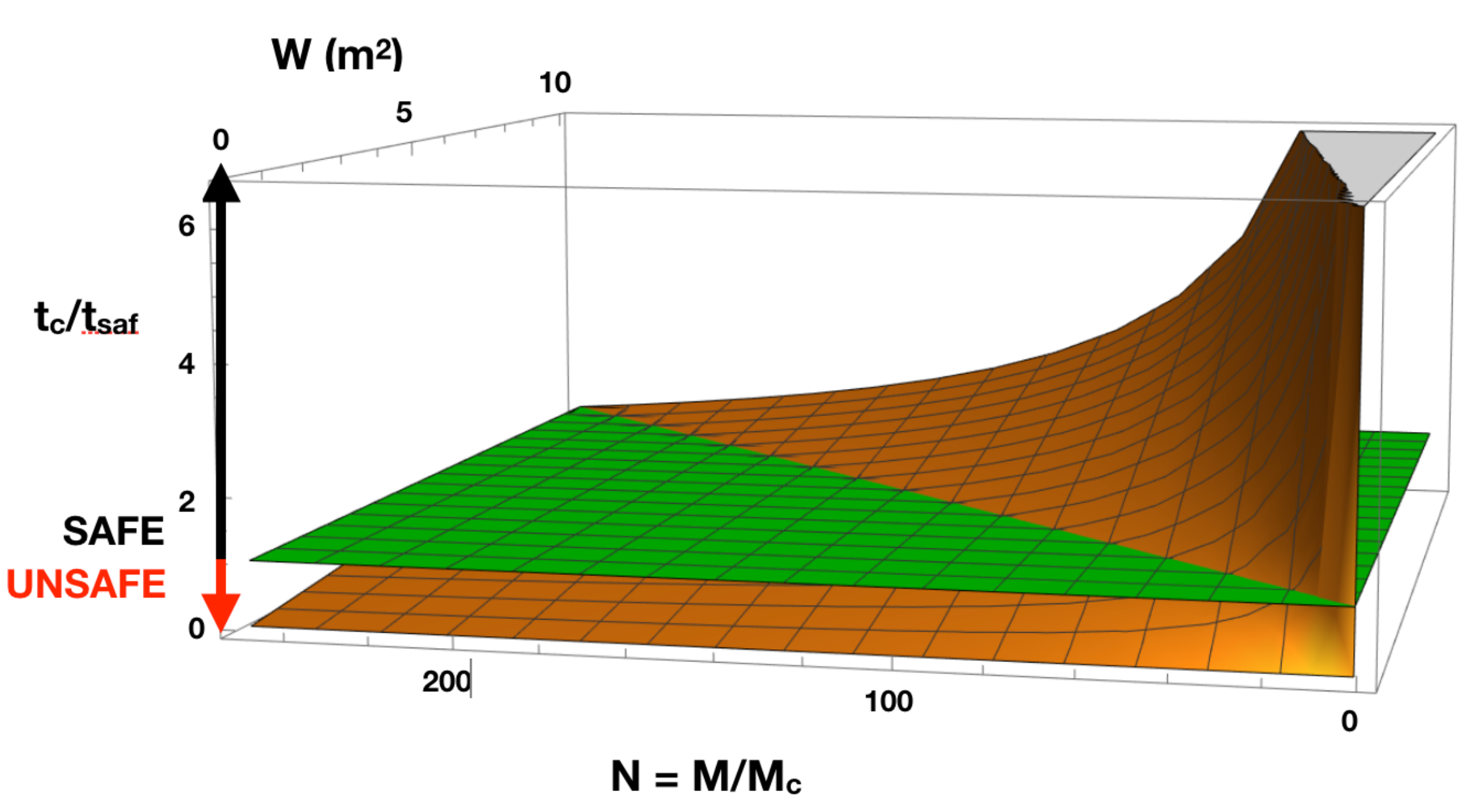}
\end{center}
\caption{ { Critical time for infection and condition for safety.} \textit{Upper panel}: the  time $t_c$  for a non-ventilated room with reflecting walls according to Eq.~(\ref{5}) plotted  as a function of the number of people that can be infected by the injected load and room radius $R$. Infection is prevented by spending a time $ t \leq t_c$.  \textit{Lower panel}: mitigation by introducing ventilation via an open window of area $W$. The planar surface is the critical surface $t_c/t_{\rm saf}=1$ above which the time scale for infection $t_c$ exceeds the time scale $t_{\rm saf}$  for which enough virons have diffused through the window.  Points on the curved surface above the  plane indicate safe occupancy duration. The line of intersection between the surfaces shows the window area needed as a function of  the viral mass injected into the room according to Eq.~(\ref{12}).}
\label{fig1}
\end{figure}

\section*{Discussion}
\label{s4}

Turbulence is hard to avoid in interior  spaces and 
 when the viral mass injected  into a turbulent room is sufficiently high,  a person can be infected over one turbulent diffusion time.  as per equation (\ref{3a}).    
More likely  are    spaces such as school classrooms or work places  where occupants remain in the room for extended periods after which virions are fully mixed  are only exposed to a critical load  over longer times. For reflecting walls and no open windows,  the   time scale for infection  is given by Eq.~(\ref{5}) and illustrated in the upper panel of  Fig. 1.
Spending more time than this  in such an unventilated  room 
would lead to infection, independent of physical distancing.  

To provide simple practical information and intuition  the role of ventilation,   we  estimated the minimum size of an open window needed to mitigate infection for a one-time viral load for the case that would otherwise cause infection on the time scale of equation (\ref{5}).  The critical window size is given by Eq.~(\ref{12}) and depends on one unknown,  the critical
VLAD mass for infection $M_{c}$.  The lower panel of  Fig. 1
 exemplifies the effect, showing that a window area $\geq 2m^2$ is enough to prevent occupants being infected for a viral load that could potentially infect 50 people. Open windows are  therefore extremely helpful.
 These conclusions are consistent with,  but complementary to  empirical approaches  such as  those of 
 of  \cite{Somsen+2020}. 
Masks reduce the area of the face and thus decrease the required open window area to obtain the same protection.

In addition to windows, retrofitting HVAC systems with UVC (ultra violet c)   or other anti-viral filters  will certainly also reduce  exposure, particularly from air passing between rooms in building complexes and depending on the 
efficiency of the filter, there is a correspondence in efficacy to  a window area of a particular size.
In recognize that   the efficacy of ventilation can be thouht of as air passing   through a window or filter of a specific area,  
note that interior walls themselves  provide a  useful large surface area. If they were imbued with anti-viral materials and  constructed to mitigate boundary layer effects, rooms could potentially be safe after one turbulent diffusion time following any VLAD injection.
Such measures are also effective for other airborne illnesses such as the flu and  offer the economic benefits of  reducing sick days, even non-pandemic circumstances.  

Precision is not required for our key results and message, 
but  its further pursuit   involves more complex  numerical modeling.  These can include turbulent eddy and droplet size spectra with size-dependent airborne survival times; different droplet-eddy coupling times as a function of droplet and eddy sizes and humidity; repeated, continuous, or time-dependent injection of VLADs from different room-locations and  sources \cite{Buonanno+2020,Stadnytskyi+2020}; temperature gradients \cite{Elperin+1996};  room geometry, inhomogeneous turbulence; and wall boundary layer effects  \cite{Lejeune+2020}.

\section*{Acknowledgments}

EB acknowldges support from NSF Grant  AST-1813298 
and 
Department of Energy grants  DE-SC0020432 and DE-SC0020434. GG acknowledges support from NSF Grant  IIS-2029095.  We thank  W.J. Forrest and J.A. Tarduno for related discussions.
None of these funding sources had any involvement or influence in any aspect of  the study.

\section*{Author Contributions}
Authors Blackman and Ghoshal jointly conceived the idea for the study. Author Blackman carried out the basic calculations and drafted the initial version of the manuscript based on joint discussions. Ghoshal further contributed to writing the manuscript,  and the interpretation  and elucidation of the results,
Corresponding author Blackman confirms  full
access to everything in this study and had final responsibility for
the decision to submit for publication.

\section*{Conflicts of Interest}
We have no conflicts of interest to report.

\section*{References}

\bibliography{covid2020ref}

\end{document}